\documentclass{amsart}
\usepackage{amsmath,amssymb,amstext}
\usepackage{mathptmx}
\usepackage[theme=default-plain,charsperline=100]{jlcode}
\usepackage[T1]{fontenc}
\usepackage{microtype}
\usepackage{pgfplots}
\usepgfplotslibrary{groupplots}
\pgfplotsset{every axis/.append style={thick, xmode=log, ymode=log, xlabel=$t$, width=7cm}}
\usepackage{algorithm}
\usepackage{xcolor}
\definecolor{plotlyblue}{RGB}{0,154,250}
\definecolor{plotlyorange}{RGB}{222,111,71}
\definecolor{plotlygreen}{RGB}{62,164,78}
\pgfplotscreateplotcyclelist{mylist}{
{plotlyblue},
{plotlyorange,densely dashed},
{plotlygreen,densely dotted},
}
\usepackage[backend=biber,style=authoryear,maxnames=99,maxcitenames=2]{biblatex}
\usepackage[margin=1in]{geometry}

\usepackage{hyperref}
\hypersetup{
	colorlinks=true,
	urlcolor=plotlyblue,
	linkcolor=plotlyorange,
	anchorcolor=plotlyorange,
	citecolor=plotlygreen
	}
\usepackage{cleveref}

\title{A note on the standard diffusion curve of TAP analysis}
\author{Toby Isaac (\href{mailto:tisaac@anl.gov}{tisaac@anl.gov})}

\begin{document}

\maketitle

In TAP reactor analysis, the \emph{standard diffusion curve} \parencite{Gleaves1997} describes the outlet flux
intensity of an inert gas that transports through a uniform 1D reactor by
Knudsen diffusion after an instantaneous pulse at the reactor's inlet. If
\(C(x,t)\) is the solution of the initial boundary value PDE,
\begin{equation*}
\begin{aligned}
  \partial_t C &= \partial_x^2 C, &x\in(0,1), t > 0,& &\text{(only diffusion in the reactor)},\\
  \partial_x C &= 0, &x=0, t > 0,& &\text{(no flux at inlet after the pulse)},\\
  C &= 0, &x = 1, t > 0,& &\text{(vacuum at outlet)},\\
  C &= \delta(x), &t = 0,& &\text{(instantaneous pulse at inlet)},
\end{aligned}
\end{equation*}
then the standard diffusion curve is
\begin{equation*}
s(t) := - \partial_x C(x,t)|_{x=1}.
\end{equation*}

Using a Fourier series expansion of \(\delta(x)\) one can show that, for \(t > 0\),
\begin{equation}
s(t) = \pi \sum_{n=0}^\infty (-1)^n (2n + 1) \exp( - (n + 1/2)^2 \pi^2 t), \label{equation:index:sdc-direct}
\end{equation}
which is the form of \(s\) that appears frequently in publications 
\parencite{Yablonsky2003,Zheng2009,Kunz2020}.

For each fixed \(t\) the series converges absolutely, and for \(t > 0.1\) it is
observed that only two terms from the sum are required for an approximation
with at most \(2.5\%\) error \parencite{Phanawadee1997}.

\begin{figure}[h]
	\caption{Using direct summation (\cref{equation:index:sdc-direct})}
	\label{fig:one}
	\begin{tikzpicture}
		\begin{groupplot}[
				group style={
					group size=2 by 2,
					xlabels at=edge bottom,
					xticklabels at=edge bottom
				},
				cycle list name=mylist
			]

			\nextgroupplot[title={(a) terms used to compute $\hat{s}$}]
				\addplot +[no markers] table [x=t, y=bigfloat, col sep=comma] {fig-1-a.csv};
				\addplot +[no markers] table [x=t, y=double, col sep=comma] {fig-1-a.csv};
				\addplot +[no markers] table [x=t, y=single, col sep=comma] {fig-1-a.csv};
			\nextgroupplot[title={(b) condition number, $\sum | s_n | / | \sum s_n |$}]
				\addplot +[no markers] table [x=t, y=k, col sep=comma] {fig-1-b.csv};
			\nextgroupplot[title={(c) relative error, $|\hat{s} - s| / |s|$}, cycle list shift=1]
				\addplot +[no markers] table [x=t, y=double, col sep=comma] {fig-1-c.csv};
				\addplot +[no markers] table [x=t, y=single, col sep=comma] {fig-1-c.csv};
			\nextgroupplot[title={(d) $\hat{s}$ (log scale)},
			legend entries={\jlinl{BigFloat}, \jlinl{Float64}, \jlinl{Float32}},
			]
				\addplot +[no markers] table [x=t, y=bigfloat, col sep=comma] {fig-1-d.csv};
				\addplot +[no markers] table [x=t, y=double, col sep=comma] {fig-1-d.csv};
				\addplot +[no markers] table [x=t, y=single, col sep=comma] {fig-1-d.csv};
		\end{groupplot}
	\end{tikzpicture}
\end{figure}
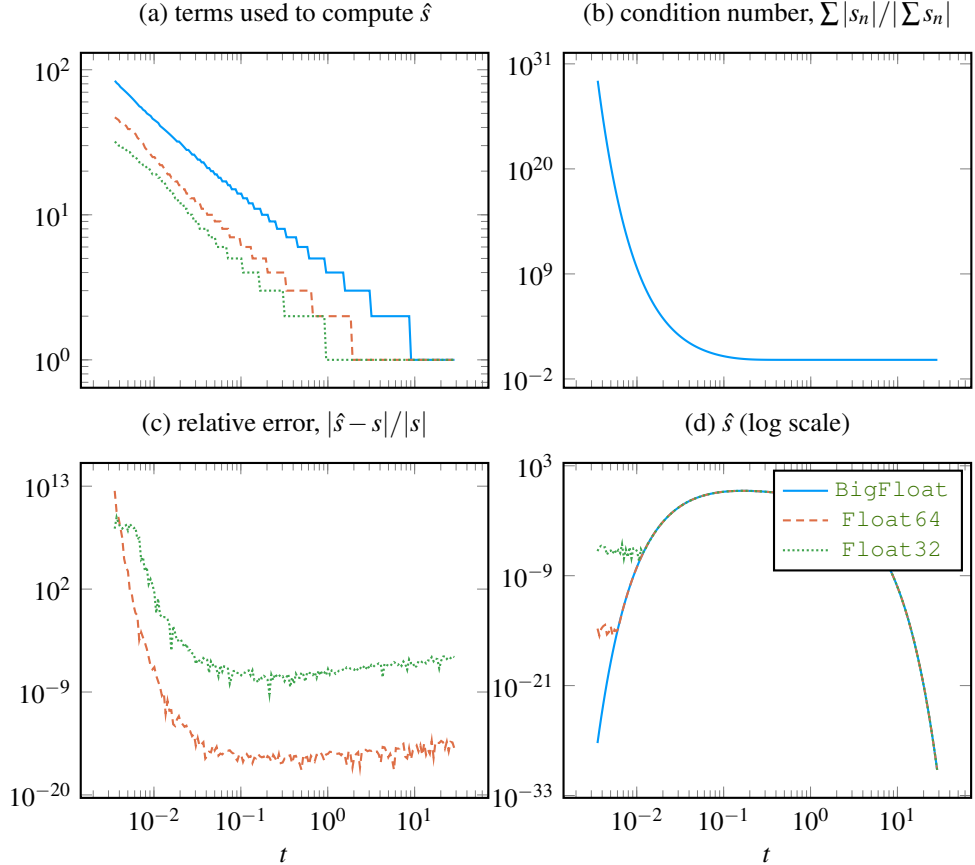

But what if you want to compute \(s(t)\) for small values of \(t\) close to
zero?  This isn't often necessary when comparing \(s\) to experimental data,
but is useful when verifying numerical TAP simulation software
\parencite{Yonge2021}.  Directly using a partial sum of
\eqref{equation:index:sdc-direct} is bad for two related reasons:
\begin{enumerate}
\item
In exact arithmetic, the number of terms required to approximate \(s\) to
a fixed relative accuracy, \(|s(t) - \hat{s}(t)| / |s(t)| < \epsilon\), is
inversely proportional to \(t\) (\cref{fig:one}(a)).

\item
The relative error of a series \(\sum_n s_n\) computed using floating point
arithmetic grows like the condition number of the sequence, \(\sum_n |s_n|
/ | \sum_n s_n|\).  For the standard diffusion curve, this quantity grows
extremely quickly as \(t \to 0\) (\cref{fig:one}(b)).  In
double precision arithmetic, the computed value will have no digits of
accuracy for \(t < 0.006\) (\cref{fig:one}(c,d)), and the
computed value may even have the wrong sign.

\end{enumerate}

We demonstrate these shortcomings with the approximation \(\hat{s}\) computed in
different floating point systems, using as many terms of the infinite sum as
are necessary for the floating point value to stabilize.

\begin{algorithm}
\begin{jllisting}
## Compute the standard diffusion curve by adding terms until the floating
#  point value doesn't change.
#
# - set abs=true to compute \sum | s_n |
# - set infty=k for only k terms
function sdc_direct(t::T; abs=false, infty=typemax(Int64))::Tuple{T,Int64} where T <: AbstractFloat
    tau = T(pi)^2 * t / 4
    s = zero(t)
    s_old = copy(s)
    sign = abs ? 1 : -1
    for n in 0:infy
        s_old = s
        s += sign^n * (2n + 1) * exp(-(2n + 1)^2 * tau)
        if (s_old == s)
            return (s * pi, n)
        end
    end
    return (s * pi, infty + 1)
end;
\end{jllisting}
	\caption{\Cref{equation:index:sdc-direct} in julia}
\end{algorithm}

As a stand-in for the true value of \(s\) we will use the same algorithm but with
julia's \jlinl{BigFloat} (an interface for \href{https://www.mpfr.org/}{GNU MPFR} \parencite{Fousse2007}) with
\(2^{-256}\) precision arithmetic, capable of \textasciitilde{}77 digits of relative accuracy.

To solve this problem, we use a remarkable functional equation satisfied by the standard diffusion curve,
\begin{equation}
(\pi t)^{3/2} s(t) = s((\pi^2 t)^{-1}), \label{equation:index:sdc-indirect}
\end{equation}
which can be proved using the \href{https://en.wikipedia.org/wiki/Poisson\_summation\_formula}{Poisson summation
formula} and various
\href{https://en.wikipedia.org/wiki/Fourier\_transform\#Tables\_of\_important\_Fourier\_transforms}{Fourier transform
identities}.
This means that to evaluate \(s(t)\) for \(t < \pi^{-1}\) (where direct summation
is unstable), we can evaluate the summation \(s(\hat{t})\) for \(\hat{t} = (\pi^2
t)^{-1} > \pi^{-1}\) (where direct summation is stable).

\begin{algorithm}[h]
\begin{jllisting}
## Compute the standard diffusion curve by eq. (2)
function sdc(t::T; infty=typemax(Int64))::Tuple{T,Int64} where T <: AbstractFloat
    t_hat = 1 / (T(pi)^2 * t)
    if isinf(t_hat)
        return (zero(t), 0)
    end
    if t > t_hat
        return sdc_direct(t, infty=infty)
    else
        s_prime = sdc_direct(t_hat, infty=infty)
        return (s_prime[1] / (t * T(pi))^(3/2), s_prime[2])
    end
end;
\end{jllisting}
\caption{\Cref{equation:index:sdc-indirect} in julia}
\end{algorithm}

\pagebreak

Using this approach:
\begin{itemize}
\item
In exact arithmetic, only the first term of the sum is required to
approximate \(s(t)\) to \(<0.6\%\) relative error for all \(t\), and only two terms
are required for \(< 4\cdot10^{-6}\%\) relative error for all \(t\) (fig. 2(a)).

\item
In floating point arithmetic, at most four terms are necessary for the sum to
converge in double precision fig. 2(b)).

\end{itemize}

\begin{figure}[h]
	\caption{Using the functional equation (\cref{equation:index:sdc-indirect})}
	\label{fig:two}
	\begin{tikzpicture}
		\begin{groupplot}[
				group style={
					group size=2 by 2,
					xlabels at=edge bottom,
					xticklabels at=edge bottom
				},
				cycle list name=mylist
			]

			\nextgroupplot[title={(a) relative error with $k$ terms},
			legend entries={$k$=1, $k$=2},
			ymin=1.e-8,
			ymax=1.0,
			clip limits=false
			]
				\addplot +[no markers] table [x=t, y=oneterm, col sep=comma] {fig-2-a.csv};
				\addplot +[no markers] table [x=t, y=twoterms, col sep=comma] {fig-2-a.csv};
			\nextgroupplot[title={(b) terms used to compute $\hat{s}$}]
				\addplot +[no markers] table [x=t, y=bigfloat, col sep=comma] {fig-2-b.csv};
				\addplot +[no markers] table [x=t, y=double, col sep=comma] {fig-2-b.csv};
				\addplot +[no markers] table [x=t, y=single, col sep=comma] {fig-2-b.csv};
			\nextgroupplot[title={(c) relative error, $|\hat{s} - s| / |s|$}, cycle list shift=1]
				\addplot +[no markers] table [x=t, y=double, col sep=comma] {fig-2-c.csv};
				\addplot +[no markers] table [x=t, y=single, col sep=comma] {fig-2-c.csv};
			\nextgroupplot[title={(d) $\hat{s}$ (log scale)},
			legend entries={\jlinl{BigFloat}, \jlinl{Float64}, \jlinl{Float32}},
			]
				\addplot +[no markers] table [x=t, y=bigfloat, col sep=comma] {fig-2-d.csv};
				\addplot +[no markers] table [x=t, y=double, col sep=comma] {fig-2-d.csv};
				\addplot +[no markers] table [x=t, y=single, col sep=comma] {fig-2-d.csv};
		\end{groupplot}
	\end{tikzpicture}
\end{figure}
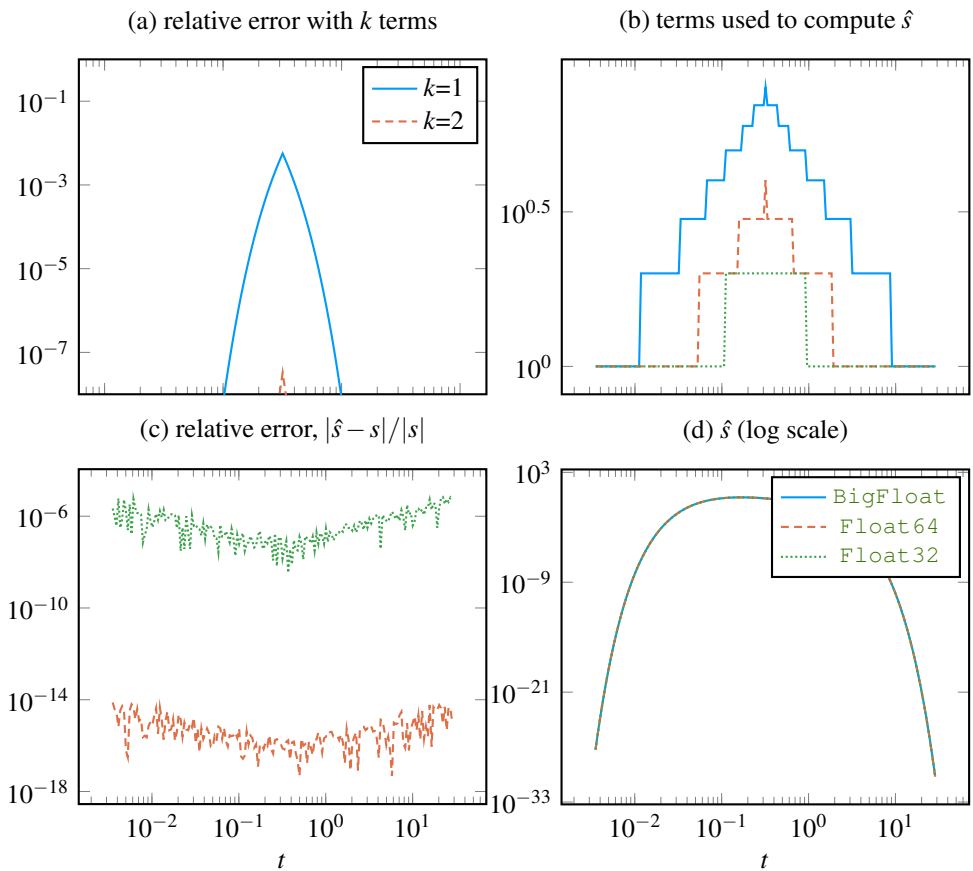

\AtNextBibliography{\footnotesize}
\printbibliography

\end{document}